# Magneto-optical evidence for a gapped Fermi surface in underdoped $YBa_2Cu_3O_{6+x}$.

L. B. Rigal

Laboratoire National Des Champs Magnétiques Pulsés, 143 Avenue de Rangueil, 31432 Toulouse, France
and Department of Physics, University of Maryland, College Park, Maryland 20742

D. C. Schmadel and H. D. Drew

Department of Physics, University of Maryland, College Park, Maryland 20742

B. Maiorov and E. Osquigil

Centro Atomico and Instituto Balseiro, S.C. de Bariloche, Argentina

J. S. Preston and R. Hughes

Department of Physics and Astronomy, McMaster University, Canada, L8S 4MI

G.D. Gu

Department of Physics, Brookhaven National Laboratory, Upton, New York 11973

The infrared (900-1100 $cm^{-1}$) Faraday rotation and circular dichroism are measured in the normal state of underdoped High $T_c$ superconductors and used to study the magneto-transport. $YBa_2Cu_3O_{6+x}$ thin films are investigated in the temperature range 10-300 K in
magnetic fields up to 8 Tesla and as a function of oxygen concentration. A dramatic increase of the Hall frequency is observed for underdoped samples which is not consistent with the approach to a Mott transition but is consistent with a partial gapping of the Fermi surface as predicted in charge density wave models.

74.25.Dw, 74.72.-h, 74.76.Bz, 78.20.Ls

The anomalous properties of the underdoped cuprates in their non-superconducting state is a major focus of attention in the efforts to understand the physics of these materials. The undoped cuprates are two dimensional Mott insulators with a large antiferromagnetic exchange interaction (J~0.2 eV). Upon hole doping the antiferromagnetism is destroyed at hole density p~0.1 and the superconductivity is optimized at p~0.16. $T_C$ decreases in the higher doped material and they appear to become conventional Fermi liquids. The underdoped metallic region between 0.1 < p < 0.16, has attracted the most attention because of anomalies in many physical properties which have led to a widespread belief that the key to understanding high temperature superconductivity lurks there. Specific heat, magnetic susceptibility, transport and optical measurements suggest a partial gapping of the Fermi surface which has been termed a pseudogap [1]. The various interpretations of this pseudogap, which include an anisotropic Mott gap or various types of CDW gaps, are controversial [1]. ARPES measurements on BSCCO show a clear Fermi surface at optimal or overdoping [2]. However, the underdoped materials show a d-wave symmetric pseudogap. The Fermi surface near ($\pi$,0) disappears leaving only "Fermi arcs" in the ($\pi$,$\pi$) direction [2]. Recently an interpretation of the Fermi arcs has been advanced in terms of a d-density wave picture [3]. In this picture charge density waves with a Q=($\pi$,$\pi$) wavevector and d-wave symmetry gaps the Fermi surface [4]. The half of the Fermi surface pocket Bragg reflected from the DDW Brillouin zone boundary is shown to have very little ARPES spectral weight. The d-symmetry in the DDW state implies a breakdown of time reversal symmetry. Charge density or spin density waves lead to a similar Fermi surface gapping. Evidence for charge ordering, or stripes, is also observed and in some cases, such as $La_{1.475}Nd_{0.4}Sr_{0.125}CuO_4$ is incontrovertible and corresponds to static charge stripes and insulation behavior [5]. At other dopings and for other materials the evidence is not so direct or unambiguous [6] and dynamic strips have been suggested.

Therefore, important questions have to do with the mode in which these systems approach the Mott state as the doping is reduced from the optimal value. In the Mott transition the Fermi surface remains intact but spectral weight transfers from a quasiparticle band at the Fermi level to the Hubbard bands away from the Fermi energy [7,8]. Equivalently, the effective mass diverges and the Fermi velocity vanishes as the Mott transition is approached. This leads to a vanishing of both the plasma frequency (i.e., Drude spectral weight) and the Hall frequency (the Hall angle spectral weight), $\omega_H \equiv eB/m_H^* c \to 0$ [8]. In stripe phase scenarios the conduction becomes pseudo one dimensional (at least locally) and the Hall conductivity is suppressed while the longitudinal conductivity is not radically affected. This leads to a large reduction in $\omega_H$ and only modest changes in the plasma frequency. On the other hand, charge or spin density waves lead to gapping, or partial gapping and breaking up of the Fermi surface into smaller pockets. In this scenario the plasma frequency reduces, because of reduced effective carrier density, and vanishes at the metal insulator transition but the Hall frequency and the cyclotron frequency, $\omega_c = 2\pi \frac{eB}{\hbar c} \left[ \oint dk_t / |v| \right]^{-1} \equiv eB/m_c c$, is enhanced as the Fermi pockets shrink. Therefore these different scenarios lead to characteristic predictions for the optical spectral weight and the Hall frequency (and cyclotron frequency) as function of carrier density. This situation makes measurement of the Hall frequency important in addressing these issues in the cuprates.

In this letter we present evidence that supports partial gapping of the Fermi surface by a measurement of the Hall frequency from high frequency magneto optical measurements on underdoped $YBa_2Cu_3O_{6+x}$. By making the measurements at frequencies above the carrier relaxation rate ($\omega\tau > 1$) the Hall frequency is obtained independently of the unknown k-space dependence of the carrier scattering rates that complicate the DC Hall effect. The experiment shows that the Hall frequency is significantly enhanced as the hole concentration is lowered below optimal doping in $YBa_2Cu_3O_{6+x}$.

The infrared Hall effect is a powerful tool for the investigation of strongly correlated metals giving important information on the carrier dynamics [9,10,11,12]. In contrast to the DC Hall effect the measurements at infrared frequencies gives information on the energy scales of current relaxation in the normal state of high Tc superconductors. It also allows tests of proposed theories which agree qualitatively with DC measurements but can be distinguished at finite frequency [9,10,11,12].

Previous measurements of the infrared Hall effect have investigated the temperature and frequency dependence of the Hall scattering $\gamma_H$ and Hall frequency $\omega_H$ for optimally doped $YBa_2Cu_3O_{6+x}$ [9,12]. From early far IR measurements (30 – 250 cm⁻ [1]) at 95 K it was concluded that $\gamma_H$ is comparable with transport scattering rate $\gamma_{xx}$ associated with longitudinal conductivity $\sigma_{xx}$[12]. More recent temperature dependent far infrared measurements (FIR) have shown an unusual square-Lorentzian spectral response, suggesting a single relaxation rate linear in temperature [10]. Measurements at mid IR frequencies (900 – 1100 cm$^{-1}$) have shown that the Hall frequency and Hall scattering rate has no discernable frequency dependence. We have repeated and improved the mid-IR measurements on optimally doped YBa$_2$Cu$_3$O$_{6+x}$ and extended them to optimally doped BSCCO and to underdoped YBa$_2$Cu$_3$O$_{6+x}$.

The measured quantity in the magneto-optical experiments is the complex Faraday angle $\theta_F = \theta_F' + i\theta_F''$, where $\theta_F'$ is the Faraday rotation and $\theta_F''$ is the circular dichroism. In our present case of low fields and strong scattering, Faraday and Hall angles are very small (a few milliradians at most at 8 T) and nearly equal to their tangents. Thus, $\theta_F \approx \tan\theta_F = t_{xy}/t_{xx}$ where $t_{xx}$ and $t_{xy}$ are the complex transmission amplitudes. Under the conditions of our experiments, $\theta_H \approx \theta_F$. However, we extract the precise $\theta_H$ from $\theta_F$ using $\sigma_{xx}$ determined separately from zero magnetic field transmittance and reflectance measurements [11].

A very sensitive technique is needed to measure such small Faraday angles. This is accomplished by modulating the IR polarization using a ZnSe photoelastic modulator (PEM) [11]. Linearly polarized IR radiation, produced by a CO$_2$ laser and a Brewster reflector, is chopped and focused on the sample in the Faraday configuration. The sample is in the center of an 8 T split coil in an optical access cryostat and the sample temperature is measured and controlled. The transmitted radiation is analyzed using the PEM and detected with a mercury-cadmium-telluride detector. Three lock-in amplifiers demodulate the resulting time-dependent signals. In this configuration, $\theta_F'$ and $\theta_F''$ are directly obtained by, respectively, the even and odd harmonic signals of the PEM modulation frequency. In this way they can be measured simultaneously with a sensitivity better than 2x10$^{-5}$ mrad/T [11].

The samples used in this study were optimally doped YBa$_2$Cu$_3$O$_7$, underdoped YBa$_2$Cu$_3$O$_{6.65}$, severely underdoped YBa$_2$Cu$_3$O$_{6.4}$ and optimally doped Bi$_2$Sr$_2$CaCu$_2$O$_8$ thin films. The YBa$_2$Cu$_3$O$_7$ samples are 100 nm thick films grown on LaSrGaO$_4$ substrates. Deoxygenation of the underdoped samples is obtained by annealing over several days at controlled temperature and oxygen pressure and cooled isobarically. The optimal and underdoped samples reported here have $T_c$'s of 89K and 55K – 60K respectively. The 6.4 sample does not show any superconducting transition down to 4.2 K. We note, also, that measurements carried out on a sample which was grown underdoped were in very good agreement with underdoped samples grown optimally doped and then annealed. The $Bi_2Sr_2CaCu_2O_8$ samples are approximately 200 nm thick films peeled from a bulk single crystal and placed on a 0.5 mm thick BaF substrate.

Before presenting our results, we note that the analysis of the Hall measurements on $YBa_2Cu_3O_{6+x}$ thin films are complicated by the Cu-O chain contributions to the conductivity. Since the longitudinal conductivity $\sigma_{xx}$ is anisotropic in single domain samples, the $\sigma_{xx}$ measured for the $YBa_2Cu_3O_{6+x}$ twinned films used in this experiment is the sum of the conductivity of the $CuO_2$ planes and, due to twinning, approximately one half of the conductivity of the chains. Since the chains do not contribute significantly to the superconductivity, it would be most interesting to focus on the transport properties of the $CuO_2$ planes by removing the chain contributions from the measurements. Because of the nearly one dimensional character of the chain bands the chain contributions to $\sigma_{xy}$ are expected to be small. In the case of single domain single crystals the chain contribution to $\sigma_{xx}$ is found to be sample dependent and comparable to the plane conductivity at 1000 cm$^{-1}$[13]. Therefore, the chain contributions cannot be reliably subtracted to obtain the pure in-plane $\sigma_{xx}$. For this reason, no chain contribution has been removed from the Hall data presented here. This leads to errors in the determination of the Cu-O plane Hall angle. However the fact that the chain conductivity is nearly frequency independent and real makes its effects on the Hall angle relatively benign and easy to characterize. The comparison of the optimally doped $YBa_2Cu_3O_{6+x}$ and chainless optimally doped BSCCO shows that while the results differ in detail the important effects are clearly observable. We will discuss the effects of the chains on the results as we discuss the data.

Figure 1 displays the results of the temperature dependence of the real part (upper panel) and the imaginary part (lower panel) of the Hall angle for the doping values of x=0.4, x=0.65 and x=0.93 at 950 cm$^{-1}$. For x=0.65 and x=0.93 doping, two different samples measured with two different techniques are shown with a very good agreement between each set of data. The imaginary part of the Hall angle has a temperature dependence quite similar for all doping, being maximum at low temperature and then decreasing in a linear-like way when the temperature increases. The real part of $\theta_H$ for 0.65 and 0.93 doping is increasing with temperature and seems to be approaching saturation around 300 K. For a severely underdoped sample (x=0.4), the trend is dramatically different with a maximum at low temperature and a slow decrease with increasing temperature. In both case, this strong temperature dependence for the Hall angle shows that contrary to $\sigma_{xx}$ which is almost temperature independent in the mid-IR, $\sigma_{xy}$ is strongly temperature dependent.

Previous frequency dependence measurements on optimally doped $YBa_2Cu_3O_{6+x}$ have shown that the spectral response function of the Hall angle in the mid-IR regime follows, approximately, a conventional Lorentzian form $\theta_H = \omega_H /(\gamma_H - i\omega)$ where $\omega_H$ is the Hall frequency (similar to the cyclotron frequency of a simple Drude metal) and $\gamma_H$ is the Hall scattering rate [9]. This Drude-like form can be obtained from Fermi liquid theory for $\gamma_H << \omega$ [9] and from many of the theoretical models for the magnetotransport in the normal state of cuprates mentioned in the introduction. With this form of spectral response, it is most revealing to plot the inverse Hall angle which should have the functional form:

$$\theta_H^{-1} = \frac{\gamma_H}{\omega_H} - i\frac{\omega}{\omega_H} \tag{1}$$

Figure 2 shows the frequency dependent measurement of $\theta_H^{-1}$ for a x=0.65 underdoped $YBa_2Cu_3O_{6+x}$ sample at 100 K and for an optimally doped $Bi_2Sr_2CaCu_2O_8$ sample at 300K. In both cases a Drude-like frequency dependence is observed with a frequency independent real part of the inverse Hall angle (upper panel) and an imaginary part linear with frequency (lower panel). While $Im(\theta_H^{-1})$ for optimally doped $Bi_2Sr_2CaCu_2O_8$ linearly extrapolates to zero the extrapolation for the underdoped YBCO film gives a positive intercept. We understand this as a consequence of the chain contributions to $\sigma_{xx}$ in YBCO from simulations with estimated chain conductances. The temperature dependence of the real part (upper panel) and the imaginary part (lower panel) of the inverse Hall angle is shown in Figure 3 for the same samples as in Figure 1.

The imaginary part depends strongly on the doping of the sample and is weakly temperature dependent while the real part of the inverse Hall angle is strongly temperature dependent. When the chain conductivity is estimated and removed from $\sigma_{xx}$ these curves are modified raising both $\mathrm{Re}(\theta_H^{-1})$ and $\mathrm{Im}(\theta_H^{-1})$ nearly uniformly compared with Fig. 3.

It is interesting to compare the real part of the mid-IR inverse Hall angle with the DC cotangent of the Hall angle. It has been found that the temperature dependence of the DC Hall angle can be written as [14]:

$$\theta_H^{-1} = b + aT^\alpha$$

where b is proportional to the impurity doping in the cuprates. While $\alpha = 2$ was reported in the earlier work, subsequent measurements on single layer and bi-layer BSCCO [15] and YBCO [16] show that $\alpha$ is oxygen doping dependent and increases as the carrier concentration decreases, from $\alpha \sim 1.75$ at optimal doping (p=0.16) to $\alpha \sim 2$ for p=0.05 underdoping. This trend in the change of the temperature dependence of the Hall angle with oxygen doping is observed in a similar but more dramatic way in our mid-IR with $\alpha \sim 1$ at optimal doping (x=0.93) and $\alpha \sim 2$ for x=0.65. This behavior even persists in the $T_c = 0$ phase (x=0.4) where $\alpha > 2$.

From the real and imaginary parts of the inverse Hall angle the temperature and frequency dependences of the Hall scattering rate and the Hall frequency can be directly extracted using Eq. 1. For all the underdoped samples, no feature in the $T^\alpha$ dependence is observed which could be linked to the opening of a pseudogap. This is in good agreement with DC Hall data which also do not show a clear anomaly at $T^*$ but the power law for $\cot(\theta_H) \sim T^\alpha$ varies with the hole density above a temperature $T_0$ which is lower than $T^*$ [17]. The same behavior is observed in the MIR range with the Hall relaxation rate power law varying across the phase diagram. In the case of underdoped samples, we cannot conclude if the temperature dependence observed in the pseudogap regime persists above $T^*$. The value of $T^*$ is not precisely known and might be above our range of temperature even for the x=0.65 sample. Moreover the characteristic temperature of the opening of the pseudogap is a crossover temperature and not a "transition" temperature. This implies a smooth temperature dependent behavior which explains the discrepancies in the determinations of T* for various experimental techniques.

For both YBa$_2$Cu$_3$O$_{6+x}$ and BSCCO samples and for all doping, little to no frequency dependence is observed for the Hall scattering rate (not shown but obtainable from Figure 2). The temperature dependence of the Hall scattering rate is shown in Figure 4 (upper panel) for optimally doped and underdoped (x=0.65) YBa$_2$Cu$_3$O$_{6+x}$ samples in the main frame and for severely underdoped YBa$_2$Cu$_3$O$_{6+x}$ (x=0.4) in the inset. The frequency quasi-independence and the strong temperature dependence of $\gamma_H$ are puzzling. In this frequency range, $\hbar\omega \gg k_b T$ over the whole range of temperature. According to both Fermi-liquid and marginal Fermi-liquid theories and as observed in angle-resolved photo emission (ARPES) [2] and optical conductivity measurements [18], the scattering rate is expected to be strongly frequency dependent and temperature independent in this range of frequencies. From the IR Hall data on both optimally doped and underdoped samples, it appears that the Hall relaxation rates remain quasi-elastic in all cases.

The lower panel of figure 4 displays the temperature dependence of the Hall frequency for the different samples. The Hall frequency is seen to increase substantially in the underdoped YBCO. It is interesting to compare these experimental values with the values that can be estimated from the ARPES data. For a Fermi liquid the Hall frequency can be expressed in terms of integrals of the Fermi velocity over the Fermi surface as [19]

$$\omega_H = \frac{eB}{\hbar c} \frac{\oint dk\, v \times dv/dk}{\oint dk |v|} = \frac{eB}{\hbar c} \frac{\oint dv \times v}{\oint dk |v|} \qquad (2)$$

where tetragonal symmetry is assumed. The second numerator is twice the area inside the Fermi velocity curve. We first compare optimally doped BSCCO for which there is the most extensive and reliable ARPES data. If it is assumed that the Fermi surface is approximately circular and the velocity varies little around the Fermi surface as is found from ARPES we find $\omega_H = (eB/\hbar c)(v_F/k_F) = 0.33\, cm^{-1}/T$, where k$_F$=0.71 $\overset{\circ}{A}^{-1}$ is the radius of the Fermi surface and v$_F$ = 1.8 eV-A, both in the zone diagonal direction [2]. The ARPES measured quantities k$_F$ and v$_F$ correspond to an effective mass, $m^* = \hbar k_F/v_F = 3 m_0$ which is about twice the band value and in good agreement with the mass deduced from the infrared conductivity. This estimation of $\omega_H$ assumes a constant velocity over the energy range corresponding to the magneto-optical experiments (100 meV). In fact the dispersion curves measured by ARPES generally show a kink at around 70meV the origin of which is controversial [2]. For optimally doped BSCCO the high frequency velocity is approximately twice the Fermi velocity. However, since the Hall frequency weights the velocity by the density of states and $D(E) \propto v(E)^{-1}$, the error introduced by using the Fermi mass for our frequency of 100 meV is less than 20%. Therefore, for optimally doped BSCCO, there is good agreement between the ARPES-deduced Fermi mass and the measured mid-infrared Hall mass while the ARPES scattering rate is very different from the Hall scattering rate.

In the case of $YBa_2Cu_3O_{6+x}$ there is comparatively little ARPES data because of the generally poor quality of the cleaved surfaces [2,20]. As a consequence, we cannot make a reliable Hall frequency comparison for $YBa_2Cu_3O_{6+x}$. However, what little data there is confirms that YBCO electronic structure, except for the chains, is very similar to that of BSCCO [18]. From ARPES data on optimally doped YBCO, we get $\omega_H = 0.22$ cm$^{-1}$ with a zone-diagonal Fermi velocity ($v_F$) of 1.3 eV-A and a Fermi surface radius of 0.74 A$^{-1}$ [20,21]. Removing the Cu-O chain contribution estimated from the literature [13] leads to an increase of about 20% of the measured mid-infrared Hall frequency which is then in good agreement with the ARPES deduced Fermi mass. Therefore, we will base our general discussion of the behavior of the Hall frequency in underdoped YBCO on the behavior of the ARPES data on underdoped BSCCO which are assumed to be similar.

For underdoped BSCCO, the ARPES dispersion curves have been measured in the $(\pi,\pi)$ direction and they show very little doping dependence of the Fermi velocity but a more rapid increase in the high frequency velocity than given by band theory [22]. Since $k_F^*$ does not change significantly with doping the Fermi mass is nearly constant or only weakly decreasing.

Therefore the observed strong increase in the Hall frequency as the hole doping is decreased in YBCO is very surprising. First, this is not the expected behavior for the approach to a Mott transition where the Hall frequency vanishes [7,8]. Also the behavior is in direct contradiction with the ARPES data on BSCCO if it is assumed that Fermi surface topology has not changed significantly, as band theory indicates, and that the zone diagonal values of the Fermi velocity are not wildly unrepresentative of the rest of the Fermi surface.

As noted earlier a strong reduction of $\omega_H$ is expected in striped phases. The measured Hall angle in the stripe phase, averaged over stripes domains, will be of the form $\theta_H \equiv \sigma_{xy}/\sigma$ where $\sigma \equiv \frac{1}{2}(\sigma_{xx} + \sigma_{yy}) \approx \frac{1}{2}\sigma_{xx}$, where x is in the (local) stripe direction. In the stripe phase $\sigma_{xy}$ (and the numerator in Eq. 2) is expected to be significantly reduced while $\sigma_{xx}$ should not be strongly affected.

Therefore, it is interesting to examine the expected behavior of the Hall frequency in the presence of a charge density wave state [3,4]. We use the DDW model although similar results are found for a CDW state. Using Eqn. 2 for the DDW energy bands the Hall frequency is found to increase substantially in the DDW state. The results for several doping levels and DDW amplitudes are presented in Table 1. The rapid onset of the increase is associated with quantum critical behavior which has been discussed for the case of the DC Hall coefficient [4]. In the calculation we have taken the tight binding parameterization of the band structure of reference [3]. $\omega_H$ is seen to increase with the DDW gap amplitude, $\Delta_0$, and decrease with hole doping, $\delta$. The changes of $\omega_H$ in the DDW state arise not so much because of the behavior of the numerator but because of the reduction in the denominator in Eq. 2. While the numerator integral is, in principle, sensitive to the curvature of the Fermi surface, in practice, the velocity curves are found to be relatively unchanged in the DDW state in the range explored. Also, from Eq. 2 the numerator integrals are seen to be relatively insensitive to the Fermi perimeter whereas, the denominator increases with the Fermi perimeter. Therefore partially gapped Fermi surfaces will generally lead to an enhancement of the Hall frequency. The enhancement of the Hall frequency also means that the Drude weight (below the DDW gap) reduces more rapidly than the Hall conductivity. Experimentally the low frequency spectral weight is observed to reduce rapidly with underdoping in the cuprates [1, 18]. This behavior corresponds, in part, to the pseudogap phenomenology within the enhanced Drude analysis of the optical data. From a comparison of the data of Fig. 4 and Table 1, the calculated Hall frequencies are seen to be in good qualitative agreement with the experimental results.

We note that no evidence for a CDW gap has been reported from IR experiments. Within the density wave scenarios this explained with the hypothesis of fluctuating density wave order. This hypothesis should not strongly affect our interpretation since the rise in Hall angle results primarily from the reduction of the $\sigma_{xx}$ spectral weight and $\sigma_{xy}$ would be affected by fluctuations only near the density wave zone boundary.

| $\delta \backslash \Delta_0$ | 0.05 meV | 0.1 meV |
|---|---|---|
| 0.07 | 2.7 | 3.7 |
| 0.10 | 2.5 | 3.5 |
| 0.16 | 2.1 | 2.9 |

Table 1: $\omega_H(\delta,\Delta_0)/\omega_H(0.16,0)$, the calculated Hall frequency at hole doping $\delta$ and d-density wave gap amplitude $\Delta_0$ normalized to the value at optimal doping ($\delta$=0.16) and $\Delta_0$=0, is given for values of $\Delta_0$ in meV and $\delta$. The calculation is based on the d-density wave model with the band parameterization of ref. 3.

In conclusion, we have reported the first measurements of mid IR Hall effect in the normal state of underdoped cuprates for different dopings. The frequency dependence of the Hall angle is Drude-like and indicates a quasi-elastic relaxation process for optimal and underdoped samples. The power law of the temperature dependence of the Hall scattering rate increases from $\alpha$ =1 to $\alpha$ >2 as the hole density is reduced from optimal doping to $T_c$ = 0. The Hall frequency is found to increase dramatically with underdoping. This large increase is inconsistent with expectations from the band structure, ARPES data, or an approach to a Mott transition in general. The rapid increase $\omega_H$ is consistent with the presence of Fermi pockets due to a partial gapping of the Fermi surface.

Further measurements are planned on underdoped BSCCO where the most extensive ARPES data exists and where the absence of chains will allow a more quantitative analysis.

We wish to thank with A. Millis, V. Yakovenko and S. Chakravarty for fruitful discussions and D. Romero for cleaving the BSCCO crystals. The research was partially supported by the NSF under Grant No. DMR-0303112

**Figure captions**

FIG. 1. Temperature dependence of the real (upper panel) and imaginary (lower panel) part of the Hall angle for $YBa_2Cu_3O_{6+x}$ thin films with x=0.93 (solid line and open circles), x=0.65 (dotted line and solid squares) and x=0.4 (dashed line).

FIG. 2. Frequency dependence of the real (upper panel) and the imaginary (lower panel) part of the Cotangent of the Hall angle for a $YBa_2Cu_3O_{6.65}$ thin film at 100K and for a $Bi_2Sr_2Cu_2O_8$ thin film at 300K.

FIG. 3. Temperature dependence of the real (upper panel) and imaginary (lower panel) part of the Inverse Hall angle for $YBa_2Cu_3O_{6+x}$ thin films with x=0.93 (solid line and open circles), x=0.65 (dotted line and solid squares) and x=0.4 (dashed line).

FIG. 4. (Upper panel) Temperature dependence of the Hall scattering rate for $YBa_2Cu_3O_{6+x}$ thin films with x=0.93 (solid line and open circles), x=0.65 (dotted line and solid squares), x=0.4 (insert) and optimally doped BSCCO(dash-dotted line). (Lower panel) Temperature dependence of the Hall frequency for $YBa_2Cu_3O_{6+x}$ thin films with x=0.93 (solid line and open circles), x=0.65 (dotted line and solid squares), x=0.4 (dashed line) and optimally doped BSCCO (dash-dotted line).

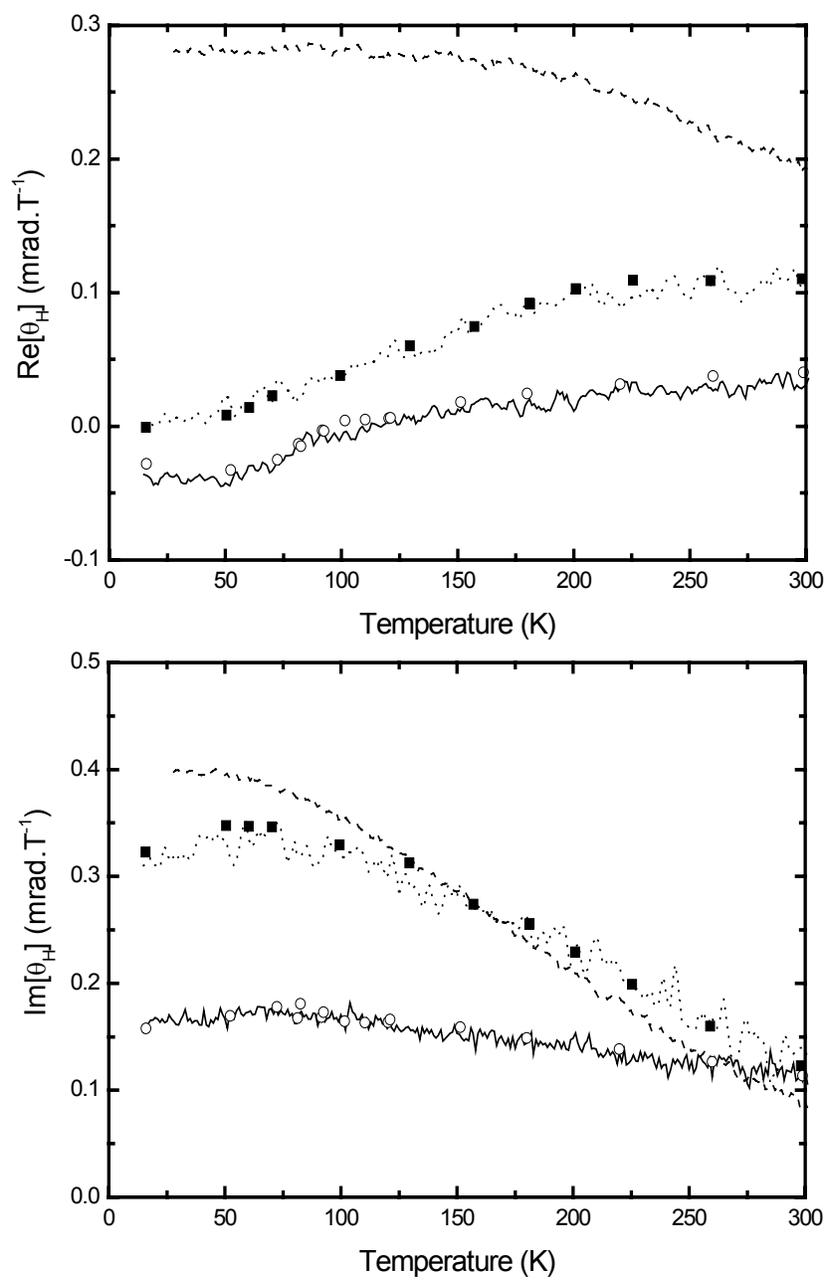

Fig.1

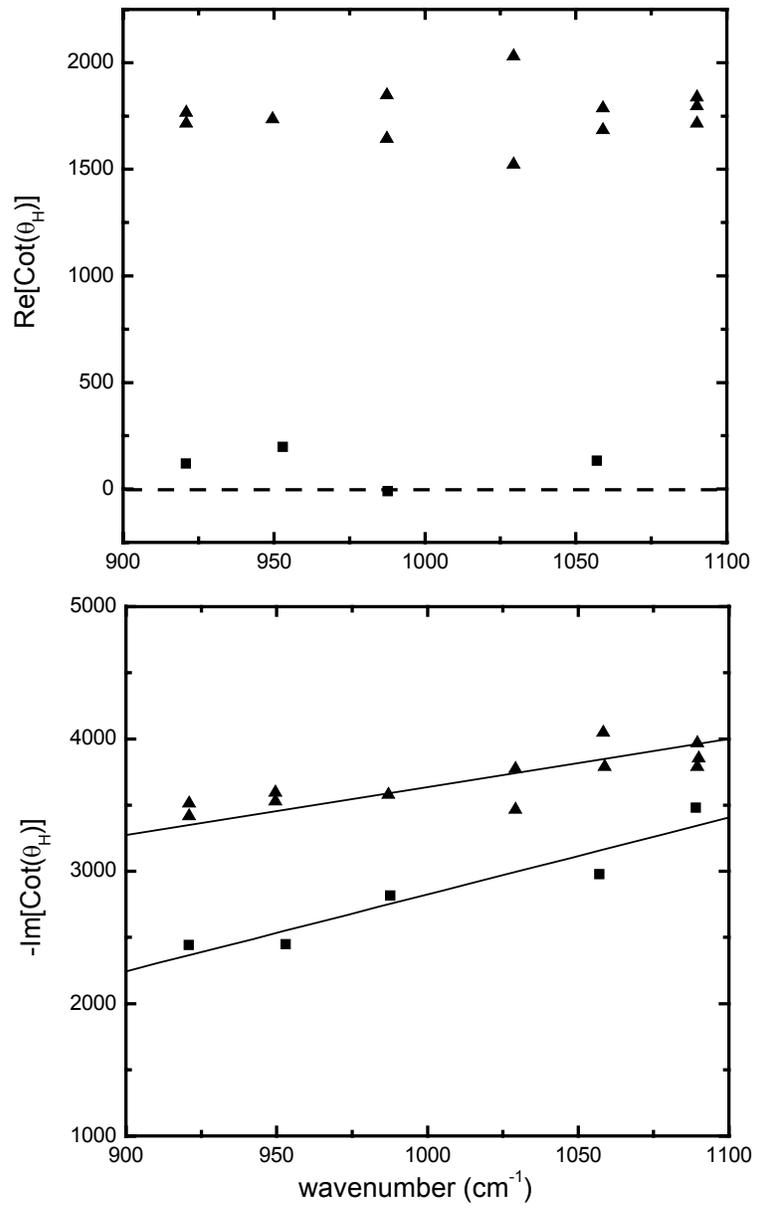

Fig. 2

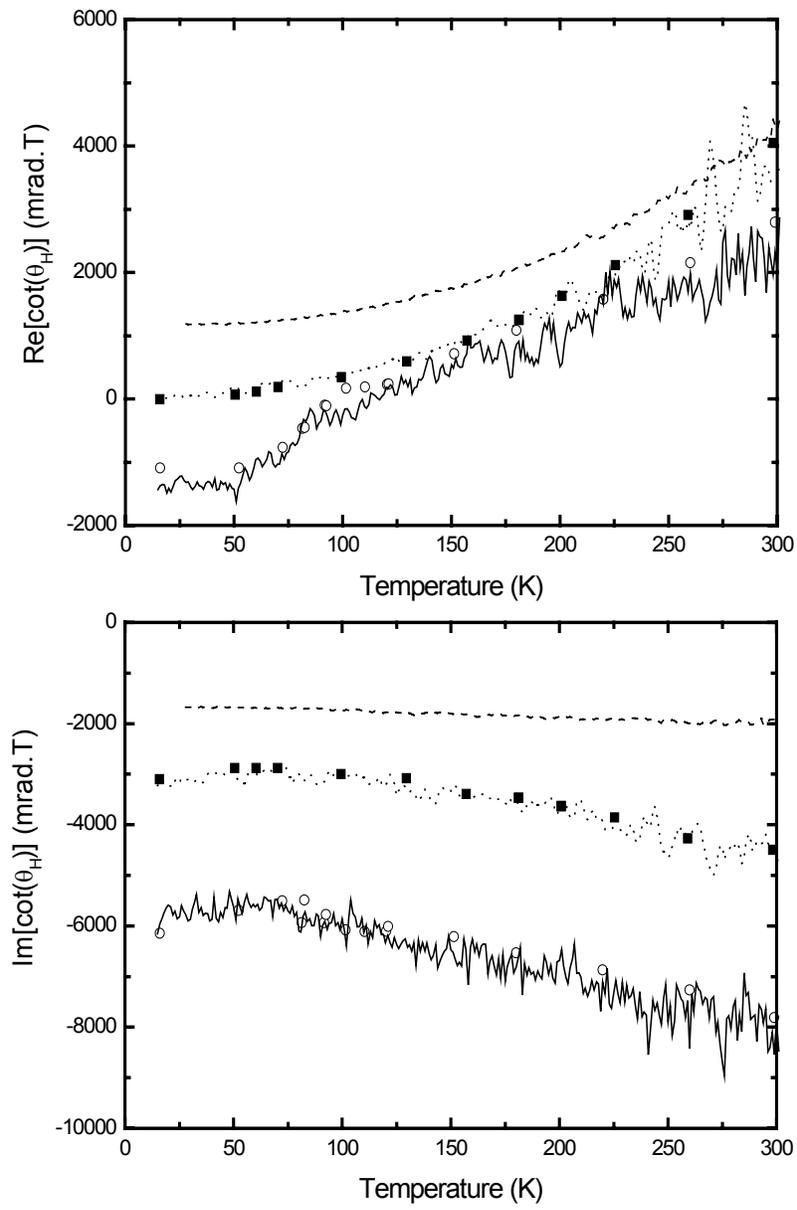

Fig. 3

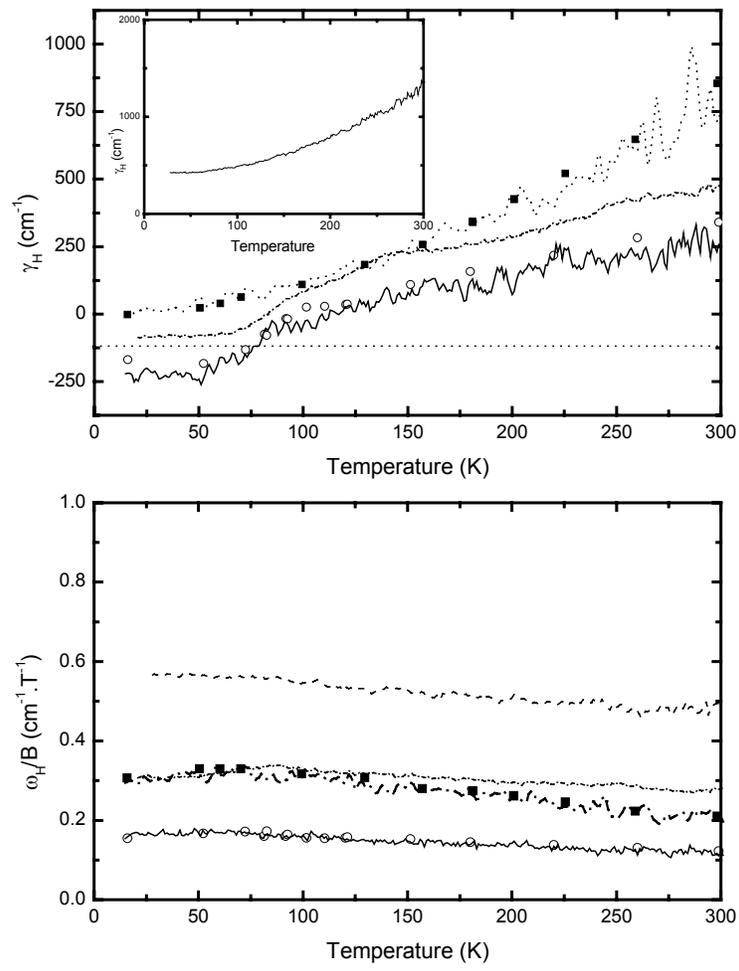

Fig. 4